\documentclass[a4paper]{spie}  

\usepackage{aas_macros}
\usepackage{amsmath,amsfonts,amssymb}
\usepackage{caption}
\usepackage{graphicx}
\usepackage[colorlinks=true, allcolors=blue]{hyperref}
\usepackage{booktabs}
\usepackage{array}
\usepackage{multirow}
\usepackage{siunitx}
\usepackage{verbatim}
\usepackage{tabularx}
\usepackage{LB}

\usepackage{lineno}

\DeclareSIUnit\Kcmb{K_{CMB}}
\DeclareSIUnit\arcmin{arcmin}

\title{\LB Mission Overview after Mission Reformation}

\input{authors_affiliations_LB_2026.12_key_spie}

\authorinfo{$^*$Corresponding emails: Tomotake Matsumura (tomotake.matsumura@ipmu.jp), Hirokazu Ishino (scishino@s.okayama-u.ac.jp)}  

\begin{document}
\maketitle

\begin{abstract}
\LB is a JAXA-led space mission designed to produce all-sky microwave polarization maps.
Its primary science goal is to test representative inflationary models by measuring the cosmic microwave background $B$-mode polarization generated by primordial gravitational waves, while also providing new insights into cosmology, particle physics, and astrophysics.
The mission concept has been updated following the reformation activities initiated after the Mission Definition Review in 2024.
The current concept preserves the central scientific objectives, while simplifying the payload configuration: a single telescope covers 12 frequency bands with band centers spanning 40 to 402\,GHz, corresponding to an optical coverage of 34--448\,GHz.
The telescope is a cross-Dragone reflector with a 500-mm aperture diameter, cooled to approximately \SI{5}{K} and coupled to transition-edge-sensor bolometer arrays operated at \SI{0.1}{K}.
\LB will observe from a Lissajous orbit around the Sun--Earth L2 point during a nominal 3-year survey.
More specifically, the primary scientific objective is to achieve total uncertainty in the tensor-to-scalar ratio of \(\delta r < 0.002\) (68\% C.L.), including contributions from foreground residuals, statistical uncertainties, instrumental systematics, and margin contingency.
The corresponding map-noise requirements are specified separately for the low-, mid-, and high-frequency ranges over the reionization and recombination multipole ranges.
This sensitivity makes \LB unique not only for inflationary science but also for a broad range of scientific investigations probing the history of both the early and late Universe, as well as for astrophysical processes, including Galactic science.
This paper summarizes the scientific objectives, updated payload and instrument concepts, observation strategy, and ground segment plans.
\end{abstract}

\section{INTRODUCTION}
\label{sec:intro}
\LB is a space mission aimed at probing primordial cosmology and fundamental physics through precision measurements of the polarization of the cosmic microwave background (CMB).  
Its primary target is the $B$-mode polarization generated by primordial gravitational waves during inflation~\cite{Kamionkowski:1996zd,Seljak:1996gy}. 
A detection of this signal would provide direct information on the energy scale and dynamics of inflation, while a stringent upper limit would exclude broad classes of inflationary models~\cite{Kamionkowski:2015yta,Komatsu:2022nvu}.  
The mission also provides a legacy all-sky, multi-frequency polarization data set for cosmology, particle physics, and astrophysics~\cite{Hazumi2021PTEP}.

\LB was selected in 2019 as the sole candidate for ISAS/JAXA's Strategic L-class Mission.  
The mission has evolved since the SPIE 2024 overview paper by Ghigna et al.~\cite{Ghigna2024}.  
After the Mission Definition Review (MDR) in 2024, the \LB team entered a reformation activity, led by Japan, with international partners Canada, France, Germany, Italy, the Netherlands, Norway, Spain, Switzerland, the United Kingdom, and the United States.
The aim of this activity was to simplify the design and preserve the science goals, while consolidating feasibility, technical responsibility, mission architecture, and procurement plans.
The updated concept replaces the previous three-telescope configuration with a single instrument and a single detection chain.
This simplification reduces payload complexity, in particular in the cryogenic chain and focal-plane interfaces, while retaining the broad frequency coverage required for component separation.

The current integrated system consists of the flight segment, the launch segment, and the ground segment. 
The flight segment comprises the payload module (PLM) and the service module (SVM).
The PLM contains the telescope, cryogenic structure, coolers, focal-plane detectors, and associated electronics.
The spacecraft is planned to be launched by JAXA's H3 launch vehicle into a transfer trajectory towards the second Sun--Earth Lagrange point (L2), where it will perform all-sky observations from a Lissajous orbit.

This paper summarizes the \LB concept established through the Mission Definition Review (MDR2), successfully completed in June 2026.
Section~\ref{sec:science_requirements} summarizes the science goals, the scientific outcomes, and the requirement flow-down.  
Section~\ref{sec:spacecraft_payload} describes the updated concept of the spacecraft and payload.  
Section~\ref{sec:operations} summarizes the planned observations and the ground-segment plans.
We conclude in Section~\ref{sec:conclusion}.

\section{SCIENCE GOAL, OUTCOMES, AND MISSION REQUIREMENTS}
\label{sec:science_requirements}

\subsection{Science goal}
The aim of the \LB mission is to create all-sky, multi-band microwave polarization maps.  These will be used to test representative inflationary models through measurements of primordial gravitational-wave $B$-mode polarization, as well as to provide new insights into cosmology, particle physics, and astrophysics.

This $B$-mode goal is motivated by the fact that scalar perturbations have been measured precisely through CMB temperature anisotropy and $E$-mode polarization observations by experiments such as ESA's \textit{Planck} satellite~\cite{Planck2018Overview}, and other CMB experiments, while tensor perturbations have not been detected yet.
Ground-based experiments including BICEP/Keck~\cite{BICEPKeck2021}, SPT~\cite{SPT3G2021}, ACT~\cite{ACTDR6_likelihood_parameters} have significantly improved sensitivity to CMB polarization and established increasingly stringent upper limits on primordial $B$ modes.
A space mission provides several unique advantages for this scientific target.
Observations from space enable stable access to the full sky, including the largest angular scales, where the primordial $B$-mode signal is expected to peak, without contamination from atmospheric emission and fluctuations. Having access to the full sky also allows us to test the isotropy of the $B$-mode signal.
In addition, broad frequency coverage across both low- and high-frequency bands is essential for accurate separation of Galactic synchrotron and thermal dust foregrounds from the faint primordial CMB polarization signal.
The primordial $B$-mode spectrum contains two important angular regimes: the reionization peak at very large angular scales; and the recombination peak around degree scales.
The combination of full-sky coverage, thermal stability, atmosphere-free observations, and wide frequency coverage makes a space mission uniquely powerful for probing primordial gravitational waves through CMB polarization.
\LB is designed to measure both $B$-mode features from space, complementing ground-based observatories, whose high sensitivity at smaller angular scales, is limited by atmospheric fluctuations and sky coverage on the largest scales.

To achieve this science goal, \LB has defined two science objectives.
The first objective is to test cosmic inflation through a search for primordial gravitational waves.
The primary observable for this objective is the tensor-to-scalar ratio \(r\), which quantifies the amplitude of tensor perturbations relative to scalar perturbations and is therefore the key parameter connecting CMB polarization observations to inflationary physics.
After the reformation process, the numerical target for the total uncertainty has been set to $\delta r < 0.002 \ (68\%~\mathrm{C.L.})$, assuming $r=0$.  
This requirement comes from a comprehensive assessment of all sources of uncertainty, including statistical noise, systematic effects, foreground residuals, lensing contributions, and contingencies and margins as allocated through the requirement flow-down.  
Correspondingly, the target for the statistical uncertainty is $\sigma_r < 0.001 \ (68\%~\mathrm{C.L.})$, including the foreground residuals and lensing contributions.  
For a fiducial value of $r=0.01$, this targeted sensitivity enables independent $3\,\sigma$  detections of both the reionization and recombination bumps. 
This target is well matched to tests of representative inflationary models, including Starobinsky-like and Higgs-inflation scenarios, at meaningful statistical significance,\cite{Starobinsky1980,Bezrukov2008} enabling \LB to probe the origin of primordial gravitational waves.\cite{Campeti2023}

\begin{figure}[tb]
    \centering
    \includegraphics[width=0.98\linewidth]{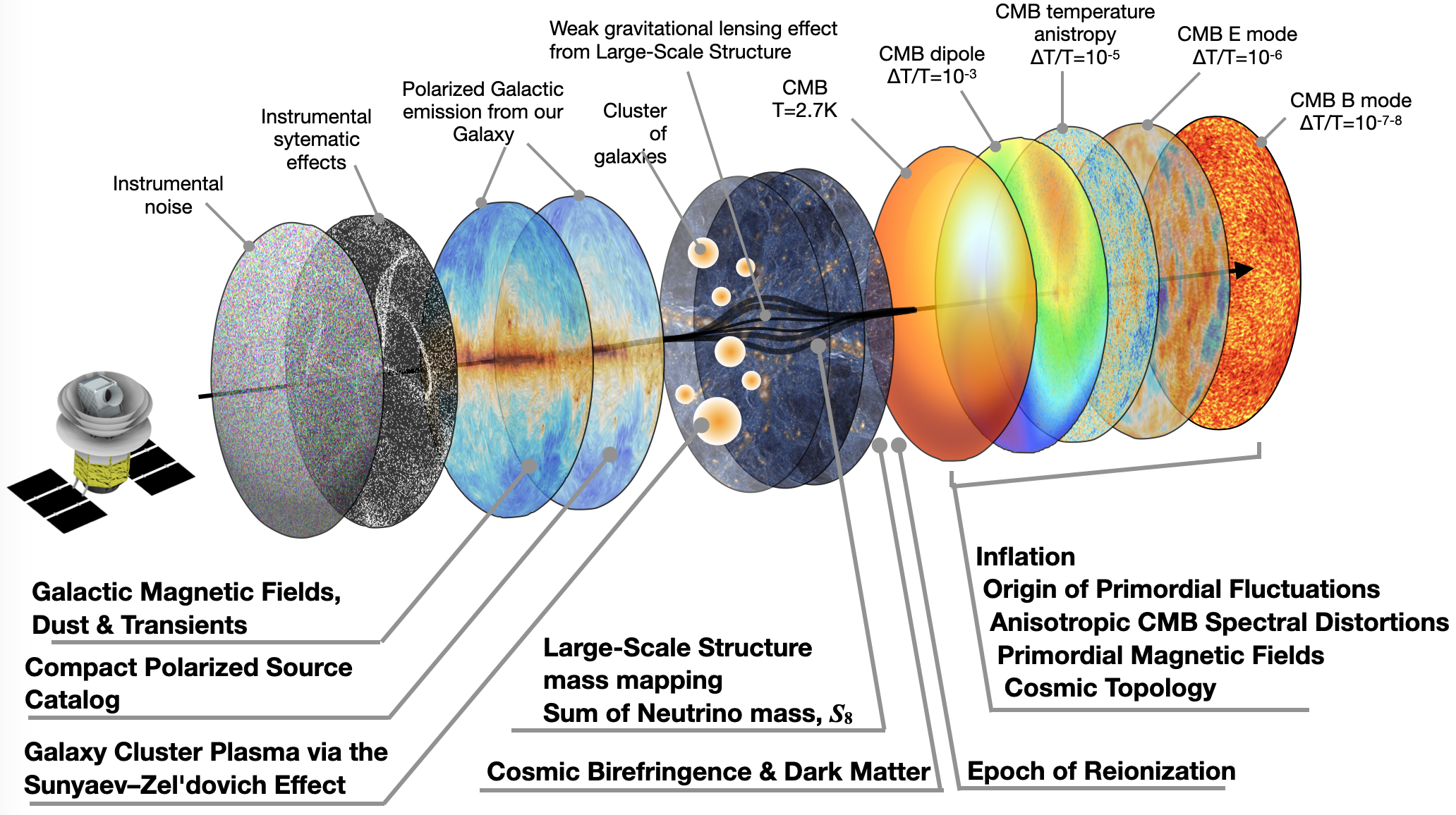}
    \caption{The \LB science objectives, probing various processes across the history of the Universe.}
    \label{fig:outcome}
\end{figure}

The second science objective is to investigate the evolution of the Universe.
\LB's high-precision full-sky multi-frequency polarized maps, defined by the first science objective, lead to a wide range of scientific outcomes, as illustrated in Fig.~\ref{fig:outcome}.

Such a set of maps will enable precise measurements of the optical depth to reionization, especially by measuring the large-angular-scale $E$ modes, which help to break degeneracies in constraining the sum of the neutrino masses.
The CMB temperature anisotropy and polarization maps can also probe `cosmic birefringence',\cite{delaHoz2025} CMB lensing,\cite{Namikawa2023,Lonoppan2023,RuizGranda2025} the Sunyaev--Zeldovich effect signals,\cite{Remazeilles2024} primordial magnetic fields,\cite{Paoletti2024} primordial non-Gaussianities, and possible anomalous statistical properties of the large-angular-scale anisotropies that could imply problems with the standard cosmological model.\cite{Banday2026} 
\LB will produce maps of the large-scale interstellar dust and Galactic magnetic field orientations, create polarization catalogs of extragalactic sources, and provide valuable data for transient and multi-messenger astrophysics. 
In combination with surveys such as ESA's \textit{Euclid}, Rubin-LSST, \textit{SPHEREx}, SKA, and ground-based CMB experiments, \LB will offer powerful cross-correlation science, probing structure formation, dark energy, baryonic physics, and galaxy evolution, establishing a long-lasting legacy data set for the broader scientific community. 

Synergy with ground-based CMB experiments remains essential. 
Space-based full-sky polarization measurements are optimized for large angular scales and foreground characterization, whereas ground-based surveys provide high angular resolution and small-scale lensing information.
Combining \LB with data from ground-based experiments, such as the Simons Observatory~\cite{SO2019}, will improve delensing and foreground modeling, and will enhance the scientific return beyond the capability of either class of experiment alone.


\subsection{Mission requirements}
The mission requirements are expressed in terms of all-sky, multi-band map products and their combined noise levels, i.e., a map-depth.  
The mission shall produce all-sky microwave polarization maps and measure multipoles from {$2 \leq \ell \leq 200$}, divided into the reionization range {$2 \leq \ell \leq 29$} and the recombination range {$30 \leq \ell \leq 200$}.  
The band centers shall span at least 40--402\,GHz, with no fewer than 12 frequency bands.  
The requirements group the bands into low-frequency~(LF), mid-frequency~(MF), and high-frequency~(HF) ranges, approximately corresponding to synchrotron-, CMB-, and dust-dominated regimes, respectively.

\begin{table}[ht]
\caption{Required combined map-noise levels for the three frequency ranges and two multipole ranges.  Units are \unit[inter-unit-product=\cdot]{\micro\Kcmb\arcmin}.}
\label{tab:map_noise_requirements}
\begin{center}
\begin{tabular}{lccc}
\toprule
Multipole range & LF: 40--100\,GHz & MF: 100--200\,GHz & HF: 200--402\,GHz \\
\midrule
\(2 \leq \ell \leq 29\)   & \(<8.0\)  & \(<10.0\) & \(<10.0\) \\
\(30 \leq \ell \leq 200\) & \(<4.7\)  & \(<4.5\)  & \(<4.7\) \\
\bottomrule
\end{tabular}
\end{center}
\end{table}

For a frequency region \(F\) and multipole range \(L\), the combined map noise is defined as
\begin{equation}
  \sigma_{F,L}=\left[\sum_{\nu \in F}\frac{\left(S^F_\nu\right)^2}{\sigma^2_{\nu,L}}\right]^{-1/2},
\end{equation}
where \(S^F_\nu\) is the frequency-weighting factor appropriate to the relevant component (i.e., the sky-averaged spectral shape of the synchrotron, CMB, and dust components) and \(\sigma_{\nu,L}\) is the band map noise inferred from the angular power spectrum of null maps.  
This requirement formulation is intentionally map-based because map noise is an observable that captures the combined performance of detector sensitivity, stability, systematic-error control, scanning, and data processing.
It is important to stress that the depth values in the map contain statistical noise, including the residuals from foreground removal, systematic effects, and margin from the potential uncertainty in the foreground model.

\begin{figure}[ht]
\centering
\hspace{-0.05\textwidth}
\begin{minipage}[l]{0.65\textwidth}
\begin{center}
    \includegraphics[width=0.98\linewidth]{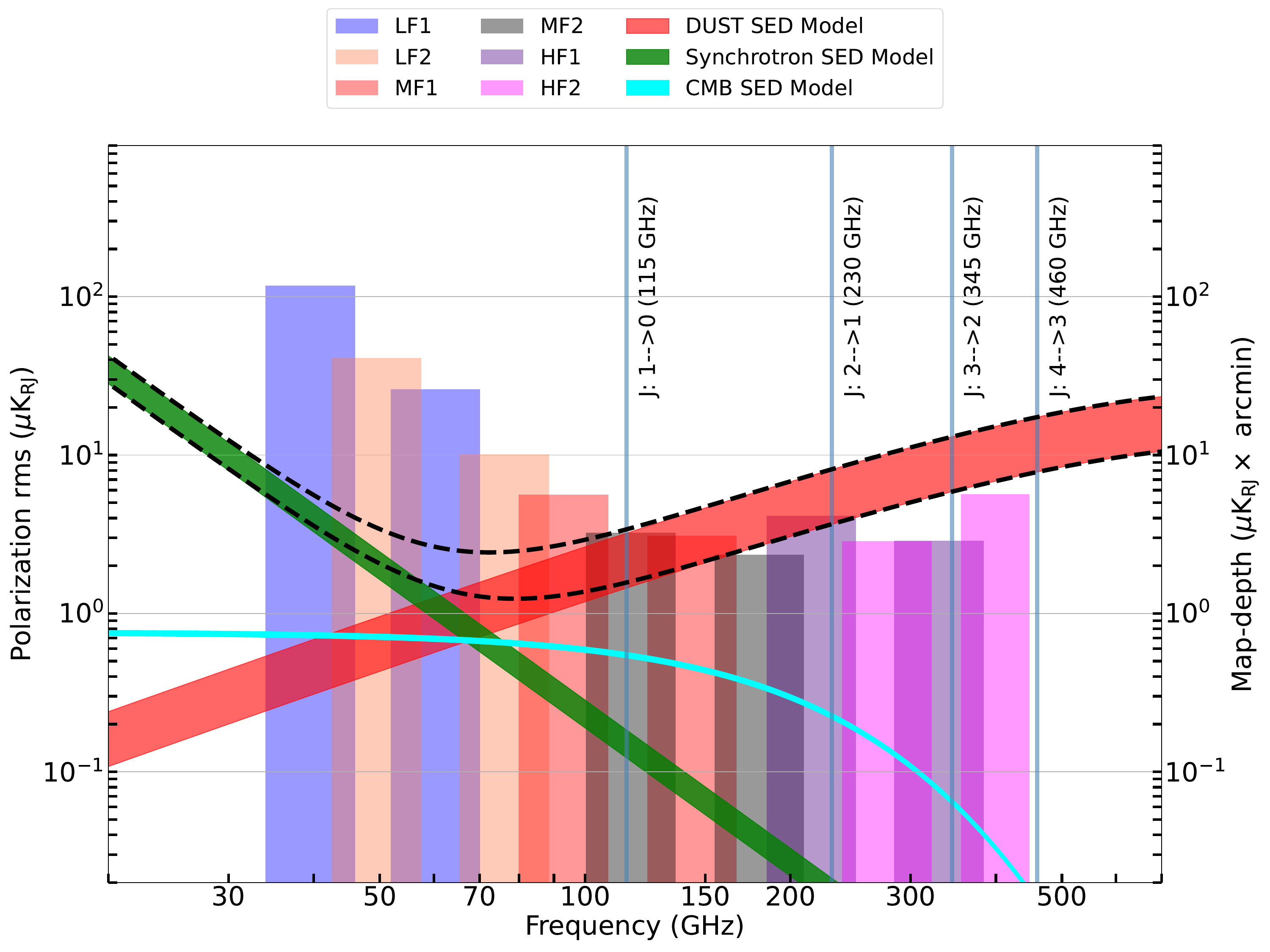}
\end{center}
\end{minipage}
\begin{minipage}[c]{0.33\textwidth}
\begin{tabular}{ccc}
\toprule
\noalign{\vskip -1pt}
Band & Center [GHz] & Range [GHz] \\
\hline\noalign{\vskip 1pt}
1    & 40           & 34.0--46.0  \\
2    & 50           & 42.5--57.5  \\
3    & 61           & 51.9--70.2  \\
4    & 77           & 65.5--88.6  \\
5    & 94           & 79.9--108.1 \\
6    & 118          & 100.3--135.7\\
7    & 145          & 123.3--166.8\\
8    & 182          & 154.7--209.3\\
9    & 217          & 184.5--249.6\\
10   & 280          & 238.0--322.0\\
11   & 334          & 283.9--384.1\\
12   & 402          & 355.8--448.2\\
\noalign{\vskip -1pt}

\bottomrule
\end{tabular}
\end{minipage}
\caption{\label{fig:frequency_bands} Proposed \LB frequency bands and sensitivities at each band, as one of the example configuration of the reformed concept. 
The synchrotron, dust, and CMB expectations are shown as green, red, and cyan bands, respectively, while the total foreground is represented by the dashed lines.
Vertical lines indicate the Galactic carbon monoxide (CO) emission frequencies.
The table lists the preliminary frequency ranges of the 12 \LB bands. }
\end{figure}

\subsection{Requirement flow-down}
The map-depth values, defined as the mission requirements, shall flow down to the system level requirements; determining this will be the central activity during Phase~A. 
The current concept design uses 12 bands.
The broad coverage (see Fig.~\ref{fig:frequency_bands}) is essential for robustly separating CMB polarization from synchrotron, thermal dust, anomalous microwave emission, and important molecular-line contributions, while also providing redundancy for internal consistency checks.
Similarly, the noise-equivalent-temperature for polarization measurements at each observational band is also derived by taking into account the feasibility of foreground removal. 
Detailed descriptions of this feasibility and the choice of mission configurations are presented elsewhere in these proceedings.
Figure~\ref{fig:frequency_bands} shows one example of the proposed frequency bands and the corresponding sensitivities. 
The exact numbers are yet to be consolidated, since this will be the primary activity during Phase~A (the Concept Design period), i.e., carrying out the design iteration to define the system configuration, starting after MDR2.

\begin{figure}[tb]
\begin{center}
    \includegraphics[width=0.99\linewidth]{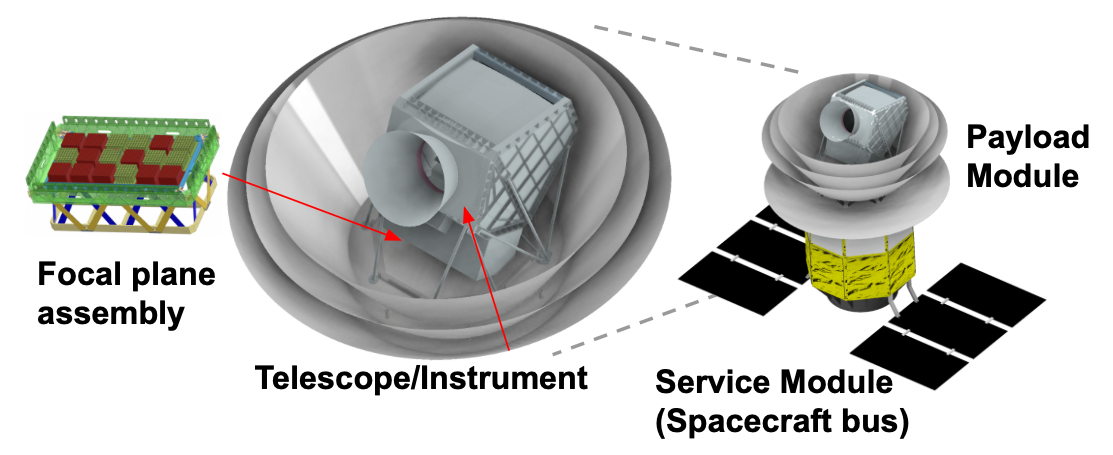}
\caption{Schematic overview of \LB.}
\label{fig:schematic_overview}
\end{center}
\end{figure}

\section{SPACECRAFT AND PAYLOAD}
\label{sec:spacecraft_payload}
The detailed mission configuration shall be frozen at System Requirement and System Definition Reviews hosted by JAXA within the Phase~A period, following MDR2.
We describe the current outlook of the mission configuration (see Table~\ref{tab:litebird_summary}), which is subject to change for further optimization. 

\subsection{Integrated system architecture}

The spacecraft consists of the PLM and SVM.  
The PLM carries the cryogenically cooled telescope and detectors; the SVM provides spacecraft bus functions, including power, telemetry, command, attitude and orbit control, and communications.  
The current design preserves an axisymmetric spacecraft philosophy suitable for continuous scanning, while adopting a simplified payload configuration relative to the 2024 three-telescope architecture, although a reflective telescope configuration inevitably breaks the axial symmetry.

\subsection{Cryogenic payload module}

The PLM cryo-structure includes three concentric V-groove radiators, a carbon-fiber-reinforced-plastic truss, mechanical coolers,  and cryogenic interface stages.  
Passive radiative cooling from the V-grooves and active mechanical cooling together reduce the temperature of the telescope structure to the few-kelvin regime.  
The baseline cooler chain uses two-stage Stirling coolers (STs) for V-groove and precooling duties, a 4-K Joule-Thomson (JT) cooler for the telescope structure, a 2-K JT cooler, and a sub-Kelvin continuous adiabatic demagnetization refrigerator~(ADR) for the focal plane.

Thermal stability is a central design driver because temperature fluctuations can couple to detector gain and optical loading.  
Current thermal analyses indicate that spin-period temperature variations at the telescope are much smaller than the relevant instrument-stability allocation, while larger but manageable variations are present on warmer V-groove stages.

\subsection{\LB Instrument}

The updated baseline instrument is a single telescope, as shown in  Fig.~\ref{fig:schematic_overview}, led by France.
It is a cross-Dragone reflecting telescope with a 500-mm aperture diameter, cooled to approximately \SI{5}{K}.
The corresponding full-width at half-maximum beamwidth ranges from \SIrange[range-phrase=\text{ to }]{6}{53}{arcmin}. 
The instrument covers the full low-, mid-, and high-frequency ranges with a single focal-plane detector unit cooled to \SI{0.1}{K}.

A cryogenic polarization modulation unit (PMU) incorporating a continuously rotating metamaterial-based transmissive half-wave plate is the baseline concept currently under study, with Italy leading its development. 
A configuration without a half-wave plate is retained as a backup option.
With a PMU, the polarization signal is modulated above the low-frequency noise knee, reducing susceptibility to \(1/f\) noise, at the cost of optical efficiency and additional loading.  
Without a PMU, polarization modulation is more strongly dependent on spacecraft spin, requiring a higher spin rate and tighter attitude-control performance. 

\subsection{Detectors}
The detector unit uses transition-edge-sensor bolometer arrays operated at \SI{0.1}{K}.
Each pixel is dichroic and senses two orthogonal polarizations.
A feedhorn and orthomode transducer (OMT) provide optical coupling to the detectors.
Each wafer is installed in a module, and 18 modules are assembled as the focal-plane unit. 
The total number of TES detectors is about 4000. 
TESs are read out via a frequency-domain multiplexing~(FDM) system, developed in Canada,\cite{Montgomery2020} combined with electronics and other hardware elements contributed by Finland, Italy, the Netherlands, and Switzerland.

One of the major outcomes of the reformation activity is that European institutes now lead detector development, with major contributions from Italy, together with the Netherlands and the United Kingdom.

\begin{table}[htbp]
\centering
\caption{Summary of the updated \LB design parameters. The polarized sensitivity is an inverse-variance weighted total.}
\label{tab:litebird_summary}
\begin{tabularx}{\linewidth}{lX}
\noalign{\vskip 5pt}
\toprule
\noalign{\vskip -1pt}
\textbf{Item} & \textbf{Specification} \\
\hline\noalign{\vskip 1pt}
Target launch &
Japanese fiscal year 2036 \\
\hline\noalign{\vskip 1pt}
Launch vehicle &
JAXA H3 rocket \\
\hline\noalign{\vskip 1pt}
Nominal observation time &
3 years \\
\hline\noalign{\vskip 1pt}
Orbit &
L2 Lissajous orbit \\
\hline\noalign{\vskip 1pt}
Satellite spin angle&
$\beta = 57.5^{\circ}$, spin rate: \SI{0.05}{rpm} with PMU \\
\hline\noalign{\vskip 1pt}
Satellite precession angle&
$\alpha = 37.5^{\circ}$, period = 1 to 10 hours \\
\hline\noalign{\vskip 1pt}
Sampling rate &
19\,Hz \\
\hline\noalign{\vskip 1pt}
Observing frequency &
34--448\,GHz \\
\hline\noalign{\vskip 1pt}
Number of distinct frequency bands &
12 \\
\hline\noalign{\vskip 1pt}
Polarization sensitivity &
\SI{2.5}{\micro\Kcmb\arcmin} (inverse noise weighted average of all bands) \\
\hline\noalign{\vskip 1pt}
Angular resolution &
53 to 6\,arcmin (FWHM of 40 to 402\,GHz) \\
\hline\noalign{\vskip 1pt}
Detectors &
Feedhorn-coupled dichroic TES bolometers (two pol.\ per pixel)  \\
\hline\noalign{\vskip 1pt}
Readout &
Frequency-domain multiplexing (FDM) \\
\hline\noalign{\vskip 1pt}
Optics &
Cross-Dragone with aperture diameter of \SI{500}{mm}, cooled to \SI{5}{K} using a combination of V-grooves, ST, and JT \\
\hline\noalign{\vskip 1pt}
Modulation &
Continuous rotating HWP polarization modulator placed at the aperture \\
\hline\noalign{\vskip 1pt}
Focal-plane temperature &
\SI{0.1}{K} with continuous ADR \\
\noalign{\vskip -1pt}
\bottomrule
\end{tabularx}
\end{table}

\subsection{Scan, Spacecraft bus, and Communications}

The spacecraft operates near the Sun--Earth L2 point in a Lissajous orbit.  
The all-sky scans combine spin, precession, and anti-Sun direction control. The 
current scan parameters in the study include a precession angle of $37.5^\circ$, a spin angle of $57.5^\circ$, a spin rate of \SI{0.05}{rpm} for the PMU option and 0.3 rpm for the non-PMU option, and a precession period between 1 and 10 hours~\cite{Takese2024}.  
The scan strategy has been optimized to maximize thermal and cryogenic feasibility within the Sun–Earth geometry constraints, while maintaining sufficient cross-linking performance, which is essential for controlling large-angular-scale polarization systematic effects.

Science data and housekeeping data are collected through the onboard network and stored in the data recorder.  
Telemetry is downlinked through an X-band link to JAXA's deep-space ground stations.  The current communication concept assumes approximately 4 hours per day of ground-station visibility, a downlink rate of the order of \SI{10}{Mbps}, and a high-gain antenna mounted on a two-axis pointing mechanism.

\section{OBSERVATIONS AND GROUND SEGMENT}
\label{sec:operations}

\LB will perform a nominal 3-year all-sky survey after the transition to L2 and completion of the commissioning phase.   
One day of observations can cover half of the sky, and the entire sky is covered after 6~months. 
Full-sky coverage from L2 gives access to the angular scales associated with reionization, which contain the modes that are difficult to measure from the ground.
Repeated full-sky observations support null tests, temporal stability checks, and systematic-error mitigation.

The ground segment is divided into the Operational Ground Segment (OGS) and Science Ground Segment (SGS) responsibilities.  
The OGS supports spacecraft integration, testing, tracking, command, telemetry reception, and mission operations.  
The spacecraft is operated from the Sagamihara Space Operation Center (SSOC) at ISAS/JAXA, with tracking by the Misasa Deep Space Station and the successor to the Uchinoura station.  

The SGS is responsible for the end-to-end processing pipeline, which delivers scientifically validated data products up to cosmological likelihoods, including simulations, calibration data ingestion, map making, component separation, and systematic error characterization.
The collaboration will produce scientific results, including scientific publications. 
The data will be released to the public by JAXA.

Calibration and verification are treated as mission-level activities.  
Calibration items include beam characterization, spectral response, instrumental polarization, polarization angle and efficiency, detector gain and linearity, transfer functions, pointing, timing, and additive noise in the optical and signal chains.  
The calibration program uses a combination of ground testing, subsystem and integrated-system verification, in-flight calibration sources, celestial sources, and end-to-end simulations.  
The purpose is not only to estimate calibration parameters, but also to confirm that residual uncertainties remain within the allocated contribution to \(\delta r\).

\section{CONCLUSIONS}
\label{sec:conclusion}

\LB is being reformed into a simpler and more consolidated mission concept, while retaining its original defining science goal: a space-based search for primordial gravitational-wave $B$-mode polarization.  
The updated architecture uses a single-telescope instrument with 12 bands having centers spanning 40--402\,GHz.
The focal-plane technology is based on feedhorn- and OMT-coupled dichroic TES bolometers cooled by ADR to \SI{0.1}{K}, which is supported by the cryogenic chain of V-groove radiators and mechanical/sub-Kelvin coolers.  
The mission requirements are expressed directly in terms of all-sky multi-band polarization maps and combined map noise levels over the reionization and recombination multipole ranges. 
The detailed mission configurations will be further optimized during the 3 years of JAXA Phase~A (Concept Design period) following the successful MDR2 in June 2026. 
After the launch, scheduled in 2036, the nominal 3-year survey from the Sun--Earth L2 region will provide both the primary inflation measurements and a broad legacy data set for cosmology, particle physics, and astrophysics.

\acknowledgments
\textit{LiteBIRD} (reformation and Phase~A) activities are supported by the following funding sources: ISAS/JAXA, MEXT, JSPS (Japan); CSA (Canada); CNES, CNRS, CEA (France); DFG (Germany); ASI, INFN, INAF (Italy); SRON (the Netherlands); RCN (Norway); MCIN/AEI, CDTI (Spain); SNSA, SRC (Sweden); UKSA (UK); and NASA (USA).

\bibliography{references}
\bibliographystyle{spiebib} 

\end{document}